\documentclass[prd,onecolumn,nofootinbib,notitlepage,10pt,aps]{revtex4-2}
\pdfoutput=1
\usepackage[utf8]{inputenc}

\usepackage{amssymb}
\usepackage[table]{xcolor}
\usepackage{graphicx}
\usepackage{psfrag,fancyhdr,epsfig}
\usepackage{hyperref}
\hypersetup{colorlinks,bookmarksopen,bookmarksnumbered,citecolor=brown,
linkcolor=blue,pdfstartview=FitH,urlcolor=blue}

\usepackage{mathtools}
\usepackage{amsmath}
\usepackage{color}
\usepackage{array}
\usepackage{graphicx}
\usepackage{color}
\usepackage{tensor}
\usepackage{xcolor}
\usepackage{orcidlink}
\usepackage{soul}

\newcommand{\be}{\begin{equation}}
\newcommand{\ee}{\end{equation}}
\newcommand{\bear}{\begin{array}}
\newcommand{\eear}{\end{array}}
\newcommand{\ba}{\begin{eqnarray}}
\newcommand{\ea}{\end{eqnarray}}

\def\a{\alpha}
\def\b{\beta}

\def\d{\delta}

\def\l{\lambda}
\def\m{\mu}
\def\n{\nu}

\def\r{\rho}
\def\s{\sigma}
\def\t{\tau}

\def\z{\zeta}

\def\tns{\tensor}

\begin{document}
\title{Dynamically induced spin-2 mass in a Weyl-invariant framework}

% \keywords{Massive spin-2 particles, Bigravity, Weyl invariance} 

\author{Ioannis D. Gialamas\orcidlink{0000-0002-2957-5276}}
\email{ioannis.gialamas@kbfi.ee}
\affiliation{Laboratory of High Energy and Computational Physics, 
National Institute of Chemical Physics and Biophysics, R{\"a}vala pst.~10, Tallinn, 10143, Estonia}

\author{Kyriakos Tamvakis\orcidlink{0009-0007-7953-9816}}
\email{tamvakis@uoi.gr}
\affiliation{Physics Department, University of Ioannina, 45110, Ioannina, Greece}

\begin{abstract}
We combine the ghost-free bimetric theory of gravity with the concept of local Weyl invariance, realized in the framework of Einstein-Cartan gravity. The gravitational sector, characterized by two independent metrics and two independent connections, is coupled to a scalar field that can in principle develop a non-vanishing expectation value through radiative corrections. The spectrum of the model, apart from the massless standard graviton and a pair of axion-like pseudoscalars, associated with the presence of the Holst invariants in the action, includes an additional spin-$2$ state of a non-vanishing Fierz-Pauli mass proportional  to the scalar field vacuum expectation value. We analyze the phenomenology of the model and specify the conditions under which the massive spin-$2$ state could be a primary dark matter candidate.
\end{abstract}

\maketitle

\section{Introduction}
\label{sec:0}
Efforts to formulate a theory of a massive spin-$2$ field have a rather long history, starting with the linear theory of Fierz and Pauli~\cite{Fierz:1939ix}. Just after the resolution of the van Dam-Veltman-Zakharov discontinuity~\cite{vanDam:1970vg,Zakharov:1970cc} issue in the framework of its non-linear generalization by Vainshtein~\cite{Vainshtein:1972sx}, the theory came to face the problem of ghosts raised by Boulware and Deser~\cite{Boulware:1972yco}. Bimetric gravity~\cite{Hassan:2011zd}, characterized by the existence of two independent metric tensors, came out as a ghost-free answer to this problem. The physical Planck mass as well as the Fierz-Pauli mass of the massive spin-2 excitation arise in terms of two independent gravitational scales, associated with the two dynamical metric tensors of the theory. On the other hand,
the problem of multiple energy scales, often vastly different, that arise in physics (hierarchy problem) has often been attributed to approximate or exact scale invariance, broken at the quantum level~\cite{Shaposhnikov:2008xb,Shaposhnikov:2008xi,Blas:2011ac,Garcia-Bellido:2011kqb,Bezrukov:2012hx,Khoze:2013uia,Karam:2015jta,Kannike:2014mia,Csaki:2014bua,Kannike:2015apa,Rinaldi:2015uvu,Ferreira:2016vsc,Marzola:2016xgb,Karananas:2016kyt,Kannike:2016wuy,Ferreira:2016wem,Ghilencea:2016dsl,Rubio:2017gty,Benisty:2018fja,Ferreira:2018qss,Karam:2018mft,Ferreira:2018itt,Gialamas:2020snr,Shaposhnikov:2018jag,Iosifidis:2018zwo,Casas:2018fum,Karananas:2021gco,Gialamas:2021enw,Cecchini:2024xoq}. For instance, the Standard model (SM) is scale invariant with the exception of the Higgs mass term. In gravitational considerations also global scale or Weyl invariance has also  been employed to address analogous issues. Weyl invariance can be promoted to a local gauge symmetry in the framework of the so called Weyl geometry. Gauge symmetries play a very important role in physics, the most obvious example being the SM of particle interactions whose success relies to a large extend on its local gauge symmetries. This philosophy can be carried over to gravitation, promoting gravity to a gauge theory of the Poincar\'e group~\cite{Utiyama:1956sy,Kibble:1961ba}. This is the Einstein-Cartan theory of gravity~\cite{Cartan:1923zea,Cartan:1924yea}, positing, in addition to the metric, a torsion tensor as a fundamental property of spacetime. The Einstein-Cartan theory can also be combined with local Weyl symmetry with the role of the corresponding Weyl gauge field~\cite{Hehl:1976kj} played by the torsion vector.

In the present article we have combined bimetric gravity (see~\cite{Schmidt-May:2015vnx} for a review and~\cite{Gialamas:2023aim,Rahman:2023swz,Gialamas:2023lxj,Hu:2023yjn,Gialamas:2023fly,Inagaki:2024ltt,Dwivedi:2024okk}~for recent applications), characterized by an additional gravitational scale that leads to a massive spin-$2$ field, with the concept of local Weyl symmetry, the latter realized in the framework of Einstein-Cartan gravity. Einstein-Cartan bigravity, in addition to the two independent metrics, posits two independent connections as well. The gravitational Weyl-invariant action (see~\cite{Edery:2014nha,Edery:2015wha,Ghilencea:2018dqd,Ghilencea:2020piz,Ghilencea:2021jjl,Ghilencea:2021lpa,Shtanov:2023lci,Wang:2022ojc,Yang:2022icz,Burikham:2023bil,Condeescu:2023izl,Karananas:2024xja,Harko:2024fnt,Condeescu:2024cjh,Glavan:2024svx} for applications) is restricted to include at most quadratic curvature invariants, so that no unphysical degrees of freedom arise, consisting of quadratic terms of the Ricci scalar curvature and Holst invariants~\cite{Holst:1995pc} corresponding to both metrics and connections. The interaction potential of the two gravitational sectors is the standard bigravity potential rendered Weyl-invariant in terms of its coupling to a scalar field. The equivalent metric form of the action features the torsion vectors corresponding to the two sectors which are non-dynamical and are integrated out. The model, apart from the two interacting spin-2 states, describes a pair of axion-like pseudoscalars, associated with the presence of the Holst terms, and the above scalar that couples to the interaction potential~\cite{Hassan:2011zd}. Considering the linear limit of the theory for proportional backgrounds, we arrive, as in the case of standard bigravity, at one massless graviton state as well as a spin-$2$ state of non-zero Fierz-Pauli mass. The latter is proportional to the vacuum expectation value (vev) of the scalar field. At this point it is argued that quantum radiative corrections, expected to break the scale invariance in a fashion analogous to dimensional transmutation, lead to a non-vanishing expectation value of the scalar field and as a result to a non-vanishing Fierz-Pauli mass. Although the theory has started with just a set of dimensionless parameters, namely those scaling the curvature invariants and those of the interaction potential, at the end it has exchanged a number of them for the set of pseudoscalar, scalar and spin-$2$ masses. 

The paper is organized as follows: In section~\ref{sec:2} we lay out the basic elements of Einstein-Cartan gravity and local Weyl invariance. We consider the simplest Weyl-invariant gravitational action that does not lead to unphysical degrees of freedom and examine its coupling to a scalar field $\phi$. In section~\ref{sec:3} we combine the above Weyl-Einstein-Cartan framework with the ghost-free theory of bimetric gravity, characterized by two independent interacting metrics and connections. The action, featuring quadratic terms of the Ricci scalars and Holst invariants, is rewritten in terms of pairs of auxiliary and axion-like pseudoscalar fields. We derive the equivalent metric form of the action and integrate out the appearing non-dynamical torsion vectors. The resulting Einstein frame action is characterized by two gravitational scales set up by the non-dynamical auxiliary fields. In addition to the two gravitons the theory describes the scalar field $\phi$ and a pair of axion-like pseudoscalars. In subsection~\ref{sec:3a} we derive the equations of motion and argue that the scalar field $\phi$ can obtain a non-vanishing expectation value through quantum corrections. In section~\ref{sec:4} and specifically in subsection~\ref{sec:4a} we discuss a special class of solutions in which the two metrics are proportional, derive the corresponding energy-momentum tensors and analyze the conditions for their compatibility imposed by the Einstein's equations. In subsection~\ref{sec:4b} we consider perturbations around proportional metric backgrounds and analyze the corresponding linear limit, characterized by a standard massless graviton and a spin-$2$ state, having a Fierz-Pauli mass. In section~\ref{sec:5} we discuss the phenomenology of the model, examining particular cases where the spin-$2$ state is ultralight or superheavy and analyzing the possibility of this state to be a primary candidate for dark matter. Finally, in section~\ref{sec:6} we state briefly our conclusions.

\section{Framework}
\label{sec:2}
In the Einstein-Cartan formulation of gravity~\cite{Cartan:1923zea,Cartan:1924yea} translations and Lorentz transformations are promoted to gauge symmetries with the gauge fields being the tetrad $e_{ \mu}^{A}$ and the spin connection $\omega_{\,\mu}^{AB}$. The corresponding field strengths can be represented by the affine-curvature ${\cal{R}}^{ \rho}_{\,\,\sigma\mu\nu}$ and the affine-torsion $\tns{T}{^\m_\n_\r}$

\be
\tns{\mathcal{R}}{^\r_\s_\m_\n} = \partial_\m \tns{\Gamma}{^\r_\n_\s} -\partial_\n \tns{\Gamma}{^\r_\m_\s} + \tns{\Gamma}{^\r_\m_\l} \tns{\Gamma}{^\l_\n_\s} -\tns{\Gamma}{^\r_\n_\l} \tns{\Gamma}{^\l_\m_\s}\,, \qquad \tns{T}{^\m_\n_\r} = \tns{\Gamma}{^\m_\n_\r} - \tns{\Gamma}{^\m_\r_\n}\,,
\ee
in terms of the affine-connection $\tns{\Gamma}{^\m_\n_\r}=e^{\mu}_{\,A}\left(\partial_{ \nu}e_{\,\rho}^{A}+\omega_{\nu B}^{\,A}e_{ \rho}^{\,B}\right)$.

The torsion tensor can be decomposed as
\be 
T_{ \mu\nu\rho}=-\frac{1}{3}\left(g_{ \mu\nu}T_{ \rho}-g_{ \mu\rho}T_{ \nu}\right)+\frac{1}{6}\epsilon_{ \mu\nu\rho\sigma}\hat{T}^{ \sigma}+\tau_{\mu\nu\rho}\,,
\ee
in terms of a vector part $T^{ \mu}$, an axial vector part $\hat{T}^{ \mu}$, defined as
\be 
T^{ \mu}=g_{ \nu\rho}T^{\nu\mu\rho},\,\,\,\,\hat{T}^{ \mu}=\epsilon^{\mu\nu\rho\sigma}T_{ \nu\rho\sigma}\,,
\ee
and a tensorial part $\tau_{ \nu\rho\sigma}$, defined by $\tns{\tau}{^\a_\mu_\a} = \tns{\tau}{^\a_\a_\m} = \tns{\epsilon}{_\m^\n^\r^\s}\tns{\tau}{_\n_\r_\s}=0$. 
	
Global scale symmetry corresponds to invariance under the transformations
\be e_{ \mu}^{\,A}\rightarrow q^{-1}e_{ \mu}^{\,A},\,\,\,\omega_{ \mu}^{AB}\rightarrow \omega_{ \mu}^{AB}\,.\ee
This symmetry can be promoted to a local gauge symmetry ({\textit{Weyl symmetry}}) introducing a gauge field $W_{ \mu}$ transforming as
\be W_{ \mu}\rightarrow W_{ \mu}+q^{-1}(x)\partial_{ \mu}q(x)\,.\ee
Under this symmetry the torsion vector transforms in an analogous fashion, namely as 
\be T_{ \mu}\rightarrow\,T_{ \mu}-3q^{-1}(x)\partial_{ \mu}q(x)\,.\ee
It is possible to realize this local symmetry without introducing the gauge field as an additional dynamical degree of freedom. Instead, one can employ the torsion vector by setting $W_{ \mu}=\frac{1}{3}T_{ \mu}$, with the corresponding  Weyl covariant derivative defined as
\be \nabla^{(w)}_{ \mu}=\partial_{ \mu}+\frac{1}{3}T_{ \mu}\,.\ee
It turns out that, in the specific Einstein-Cartan quadratic models to be considered below, the torsion is non-dynamical and, therefore, employing the above identification will not introduce any additional dynamical degrees of freedom.

The set of (local) Weyl transformations are\footnote{ Scalar fields transform in terms of their canonical dimension as $\phi\rightarrow q^{d}\phi$.} 
\be\bear{l}
g_{ \mu\nu}\rightarrow\,q^{-2}g_{\mu\nu},\,\,\,\,\,\,\tns{\Gamma}{^\m_\n_\r}\rightarrow\,\tns{\Gamma}{^\m_\n_\r}-q^{-1}\partial_{ \nu}q \delta^{ \mu}_{\,\rho}\,,\\
\,\\
{\cal{R}}\rightarrow\,q^2{\cal{R}},\,\,\,\,\,\,\tilde{\cal{R}}\rightarrow\,q^2\tilde{\cal{R}}\,,\\
\,\\
T_{ \mu}\rightarrow T_{ \mu}-3q^{-1}\partial_{ \mu}q,\,\,\,\,\,\,\,\hat{T}_{ \mu}\rightarrow\,\hat{T}_{ \mu},\,\,\,\,\,\tau_{ \mu\nu\rho}\rightarrow\,q^{-2}\tau_{ \mu\nu\rho}\,,
\eear
\ee 
where ${\cal{R}}$ and $\tilde{\cal{R}}$ are the two existing curvature scalars, namely the {\textit{Ricci scalar}} and the {\textit{Holst invariant}}~\cite{Holst:1995pc}, defined as
   \be {\cal{R}}=g^{\sigma\nu}{\cal{R}}^{ \rho}_{\,\,\sigma\rho\nu},\,\,\,\,\,\,\,{\tilde{R}}=\epsilon^{\mu\nu\rho\sigma}{\cal{R}}_{ \mu\nu\rho\sigma}\,.\ee

Constructing a theory of pure gravity in the above described  Einstein-Cartan framework and subject to the criterion of the absence of any unphysical degrees of freedom, we are restricted to consider at most quadratic terms of the curvature and in particular at most quadratic terms of ${\cal{R}}$ and $\tilde{\cal{R}}$. In addition Weyl symmetry excludes the linear terms. Therefore, a general Weyl-Einstein Cartan pure gravity action is\footnote{We exclude the parity-odd mixed ${\cal{R}}\tilde{\cal{R}}$ term.  See~\cite{Gialamas:2024iyu} and~\cite{Karananas:2025xcv} for the inclusion of this term in the context of inflation.}
\be 
{\cal{S}}_g=\int{\rm d}^4x\sqrt{-g}\left(\frac{1}{4f^2}{\cal{R}}^2\,+\,\frac{1}{4\tilde{f}^2}\tilde{\cal{R}}^2\right){\label{ACT-0}}\,,
\ee
where $g$ is the metric determinant and $f,\tilde{f}$ are dimensionless parameters.
It is also worth mentioning that the above action coincides with the metric-affine case where non-metricity is nonzero. This equivalence arises due to a projective symmetry of the action, generated by a single vector.\footnote{A new extended projective symmetry, generated by a pair of vectors and leading to interesting phenomenology, has been discussed in~\cite{Barker:2024dhb}.} However, if we explicitly include terms involving non-metricity and/or torsion in the action (as in what follows)~\cite{Iosifidis:2021bad,Rigouzzo:2022yan,Rigouzzo:2023sbb}, this equivalence no longer holds.
 
The action~\eqref{ACT-0} can also be written in terms of two auxiliary fields, namely an {\textit{auxiliary scalar}}  $\chi$ and an {\textit{axion-like}} $\zeta$ as
\be 
{\cal{S}}_g=\int{\rm d}^4x\sqrt{-g}\left(\frac{\chi}{2f^2}{\cal{R}}+\frac{\zeta}{2\tilde{f}^2}\tilde{\cal{R}}-\frac{\chi^2}{4f^2}-\frac{\zeta^2}{4\tilde{f}^2}\right)\,.{\label{AUX}}
\ee
The coupling of a scalar field $\phi$ to gravity in this framework will necessarily restrict its self-interaction  potential to be quartic. The corresponding action will be
\be 
{\cal{S}}_{m}=\int{\rm d}^4x\sqrt{-g}\left(\frac{1}{2}\xi\phi^2{\cal{R}}-\frac{1}{2}(\nabla^{(w)}_{ \mu}\phi)^2-\frac{\lambda}{4}\phi^4+C\hat{T}^2\phi^2\right)\,.
\ee
Note also the presence of the Weyl-invariant term $\sqrt{-g}\hat{T}^2\phi^2$, where $\hat{T}^2=g^{\mu\nu}\hat{T}_{\mu}\hat{T}_{\nu}$, which cannot be excluded. A non-minimal coupling term $\sqrt{-g}\tilde{\xi}\phi^2\tilde{\cal{R}}$ is also allowed by Weyl invariance, although it is parity odd.  Recently, numerous studies have explored the effects of this term in inflation~\cite{Langvik:2020nrs,Shaposhnikov:2020gts,Salvio:2022suk,Gialamas:2022xtt,DiMarco:2023ncs,He:2024wqv,Gialamas:2024iyu,Racioppi:2024zva,Racioppi:2024pno,Gialamas:2024uar,Katsoulas:2025mcu,Karananas:2025xcv}.

Closing this section we note that the action can be set in a metric form by substituting in place of the curvature scalars the formulae
 \begin{subequations}
 {\label{METRIC-0}}
\begin{align}
{\cal{R}} &=R(g)+2\nabla_\m T^\m-\frac{2}{3}T_\m T^\m+\frac{1}{24}\hat{T}_\m \hat{T}^\m +\frac12 \t_{\m\n\r}\t^{\m\n\r}\,,
\\
\tilde{\cal{R}}&=-\nabla_\m \hat{T}^\m+\frac{2}{3}\hat{T}_\m T^\m +\frac12\epsilon^{\m\n\r\s}\t_{\l\m\n}\tns{\t}{^\l_\r_\s}\,,
 \end{align}
\end{subequations}
 where the metric Ricci scalar $R(g)$ and the covariant derivative are defined in terms of the Levi-Civita connection.  In the above formulae, the quadratic terms of the tensorial part $\tau_{ \mu\nu\rho}$ can be neglected, as their (algebraic) equation of motion yields $\tau=0$, allowing us to set them to zero to begin with. 
\section{The model}
\label{sec:3}

Bimetric gravity as a ghost-free theory of a massive spin-$2$ particle, positing two independent metric tensors $(g_1)_{\mu\nu}$ and $(g_2)_{ \mu\nu}$, formulated in the Einstein-Cartan framework, is also characterized by two independent connections. The Fierz-Pauli mass for one of the spin-$2$ states is generated by a standard interaction potential of the form\footnote{Starting with $e_0(X)=1$, the polynomials $e_n(X)$ satisfy the recurrence relation
$$e_n(X)=\frac{(-1)^{n+1}}{n}\sum_{k=0}^{n-1}(-1)^kTr(X^{n-k})e_k(X)\,.$$} $\sum_{n=0}^{4}\beta_ne_n(\sqrt{\Delta})$, where $\tns{\Delta}{^\m_\n}=g_1^{\,\mu\rho}g_{2\,\rho\nu}$, in terms of five parameters $\beta_n$.
Imposing invariance under common local transformations $g_{a\, \mu\nu}\rightarrow q^{-2}g_{a\, \mu\nu}$ and $\tns{(\Gamma_a)}{^\m_\n_\r}\rightarrow\,\tns{(\Gamma_a)}{^\m_\n_\r}-q^{-1}\partial_{ \nu}q \delta^{ \mu}_{\,\rho} $ leads us to consider an action of the form~({\ref{ACT-0}}) for each metric. Thus, for the non-interacting part of the gravitational action we adopt the action\footnote{ The impact of a quadratic curvature term in bigravity has been recently discussed in~\cite{Gialamas:2023lxj,Gialamas:2023fly} using the standard metric formulation and in~\cite{Gialamas:2023aim} within the framework of metric-affine gravity. For further applications in the context of $F(R)$ massive gravity and bigravity, refer to~\cite{Kluson:2013yaa, Nojiri:2012re}.}
\be 
{\cal{S}}_g=\int{\rm d}^4x\left[\sqrt{-g_1}\left(\frac{1}{4f_1^2}{\cal{R}}_1^2+\frac{1}{4\tilde{f}_1^2}\tilde{\cal{R}}_1^2\right)+\sqrt{-g_2}\left(\frac{1}{4f_2^2}{\cal{R}}_2^2+\frac{1}{4\tilde{f}_2^2}\tilde{\cal{R}}_2^2\right)\right]\,.
\ee
Note that the standard interaction potential discussed above is of zero Weyl weight. A Weyl-invariant interaction potential that preserves the results of bimetric gravity can be written in terms of a scalar field $\phi$ (of canonical dimension $d=1$) with a quartic interaction, namely
\be V(\sqrt{\Delta},\phi)=\phi^4\sum_{n=0}^{4}\beta_ne_n(\sqrt{\Delta})\,,
\ee
where $\beta_n$ are five dimensionless parameters.
Note that a quartic self-interaction is included in the sum in terms of the $n=0$ and $n=4$ terms.
Thus, the full action of the model is
\be {\cal{S}}={\cal{S}}_g+{\cal{S}}_{{\rm int},\phi}\,,\ee
where
\be {\cal{S}}_{{\rm int},\phi}=\int{\rm d}^4x\sqrt{-g_1}\left(-\frac{1}{2}(\nabla^{(w)}_{ \mu}\phi)^2+C\hat{T}_1^2\phi^2+\phi^4\sum_{n=0}^{4}\beta_ne_n(\sqrt{\Delta})\right)\,.\ee

The specific form of the potential $V$ is determined by the requirement that the theory remains both ghost-free~\cite{Hassan:2011zd,Hassan:2011ea} and invariant under Weyl transformations. This ensures the internal consistency of the framework, preventing the emergence of unphysical degrees of freedom while preserving its fundamental symmetry properties. Terms of the form $\xi \phi^2 R$ and $\tilde{\xi} \phi^2  \tilde{R}$ also preserve Weyl invariance. However, we omit them at this stage for the sake of simplicity.

We can proceed by introducing auxiliary fields in the fashion of~({\ref{AUX}}), namely a pair of auxiliary scalar fields $\chi_a$ and a pair of axion-like pseudoscalars $\zeta_a$, and rewrite the gravitational part of the action, summing up over the two sets of variables in terms of an index $a$, as
\be {\cal{S}}_g=\int {\rm d}^4x\sum_{a=1,2}\sqrt{-g_a}\left(\frac{\chi_a}{2f_a^2}{\cal{R}}_a+\frac{\zeta_a}{2\tilde{f}_a^2}\tilde{\cal{R}}_a-\frac{\chi_a^2}{4f_a^2}-\frac{\zeta_a^2}{4\tilde{f}_a^2}\right)\,.\ee
Weyl invariance allows us to set one of the auxiliary fields to 
\be 
 \chi_1=f_1^2 m_1^2\,.
\ee
Next, we can employ the expressions ({\ref{METRIC-0}}) in terms of the torsion vectors and write the full action in metric form as
\begin{equation}
\label{eq:act_2}
    \mathcal{S} =  \mathcal{S}_1 + \mathcal{S}_2 +  \mathcal{S}_{\rm int}\,,
\end{equation}
with
\begin{subequations}
\begin{align}
    &\mathcal{S}_1 = \int {\rm d}^4x \sqrt{-g_1} \bigg[ \frac{m_1^2}{2} \left(R[g_1]-\frac{2}{3}T_1^2 +\frac{1}{24}\hat{T}_{1\mu}\hat{T}_1^{\mu}\right) +\frac{\bar{\z}_1}{2}\left(\frac23 T_1^\m \hat{T}_{1\mu} \right) +\frac12 \hat{T}_1^\mu \partial_\mu \bar{\z_1} -\frac12 (\partial_\m \phi)^2\nonumber
\\
    & \hspace{3.2cm} -\frac{1}{18}\phi^2 T_1^\m T_{1\m} -\frac13\phi T_1^\m\partial_\m\phi +    C\phi^2 \hat{T}_1^2 -\frac{f_1^2}{4}m_1^4 - \frac{\tilde{f}_1^2}{4}\bar{\z}_1^2 \bigg]\,,
\\[0.2cm]
    &\mathcal{S}_2 = \int {\rm d}^4x \sqrt{-g_2} \bigg[ \frac{\bar{\chi}_2}{2} \left(R[g_2]-\frac{2}{3}T_2^2 +\frac{1}{24}\hat{T}_{2\mu}\hat{T}_2^{\mu}\right) +\frac{\bar{\z}_2}{2}\left(\frac23 T_2^\m \hat{T}_{2\mu}  \right)  +\frac{1}{2}\hat{T}_2^\m \partial^\m \bar{\zeta}_2 - T_2^\m \partial \bar{\chi}_2\nonumber
    \\
    & \hspace{3.2cm} - \frac{f_2^2}{4}\bar{\chi}_2^2 - \frac{\tilde{f}_2^2}{4}\bar{\z}_2^2 \bigg]\,,
\\[0.2cm]
&\mathcal{S}_{\rm int} =  \int {\rm d}^4x \sqrt{-g_1}  \phi^4\sum_{n=0}^{4}\b_n e_n(\sqrt{\Delta})\,. 
\end{align} {\label{METRIC-1}}
\end{subequations}
Note that the kinetic and $\hat{T}^2$ terms of $\phi$, included in ${\cal{S}}_{int,\phi}$ are not included in ${\cal{S}}_{int}$, having been incorporated in ${\cal{S}}_1$. Note also that we have omitted the terms $\tau_1^{\m\n\r}, \tau_2^{\m\n\r}$ since they appear quadratically in the action, and their equations of motion yield $\tau_1^{\m\n\r}=\tau_2^{\m\n\r}=0$ as solutions. We have imposed the gauge $\chi_1 = f_1^2m_1^2$ and redefined the fields as $\z_1/\tilde{f}_1^2 =\bar{\z}_1,\, \z_2/\tilde{f}_2^2 =\bar{\z}_2$ and $\chi_2/f_2^2 =\bar{\chi}_2$.

Varying the above action~({\ref{METRIC-1}}) with respect to $T_1, \hat{T}_1, T_2$ and $\hat{T}_2$ we obtain the system of linear equations with solutions 

\begin{align}
    T_1^\m & =- \frac{3\left[12\bar{\zeta}_1 \partial^\m \bar{\zeta}_1 + \phi \partial^\m\phi(m_1^2+48C \phi^2)\right]}{24\bar{\zeta}_1^2 + (6m_1^2+\phi^2)(m_1^2+48C\phi^2)}\,, \quad  \hat{T}_1^\m  = \frac{12\left[2\bar{\zeta}_1\phi \partial^\m \phi - \partial^\mu \bar{\zeta}_1 (6m_1^2 +\phi^2)\right]}{24\bar{\zeta}_1^2 + (6m_1^2+\phi^2)(m_1^2+48C\phi^2)}\,, \nonumber
\\[0.2cm]
    T_2^\m & = - \frac{3(4\bar{\zeta}_2\partial^\m \bar{\zeta}_2+\bar{\chi}_2 \partial^\mu \bar{\chi}_2)}{2(\bar{\chi}_2^2+4\bar{\zeta}_2^2)}\,, \hspace{2.72cm} \hat{T}_2^\m  = \frac{12(\bar{\zeta}_2 \partial^\m \bar{\chi}_2 - \bar{\chi}_2 \partial^\mu \bar{\zeta}_2)}{\bar{\chi}_2^2+4\bar{\zeta}_2^2}\,.
\end{align}

Substituting the solution back into the action~\eqref{eq:act_2} we obtain
\begin{subequations}
\begin{align}
    &\mathcal{S}_1 = \int {\rm d}^4x \sqrt{-g_1} \bigg[ \frac{m_1^2}{2} R[g_1]  -\frac{f_1^2}{4}m_1^4 - \frac{\tilde{f}_1^2}{4}\bar{\z}_1^2
    \\
    & \hspace{3.cm} -3\frac{(\partial_\m \bar{\z}_1)^2 (6m_1^2+\phi^2) + (\partial_\m \phi)^2(m_1^4+4\bar{\z}_1^2+48Cm_1^2\phi^2) - 4\bar{\z}_1\phi \partial_\m\phi\partial^\m\bar{\z}_1}{24\bar{\z}_1^2 +(6m_1^2+\phi^2)(m_1^2+48C\phi^2)}   \bigg]\,, \nonumber
\\[0.2cm]
    &\mathcal{S}_2 = \int {\rm d}^4x \sqrt{-g_2} \left[ \frac{\bar{\chi}_2}{2} R[g_2]  - \frac{f_2^2}{4}\bar{\chi}_2^2 - \frac{\tilde{f}_2^2}{4}\bar{\z}_2^2 +\frac{3}{4}\left(\frac{\bar{\chi}_2 (\partial_\m \bar{\chi}_2)^2-4\bar{\chi}_2 (\partial_\m\bar{\z}_2)^2+8 \bar{\z}_2\partial_\m\bar{\chi}_2 \partial^\m\bar{\z}_2}{\bar{\chi}_2^2 +4\bar{\z}_2^2}\right)\right]\,.
\end{align}
\end{subequations}
The action $\mathcal{S}_1$  can be diagonalized by performing the field redefinition~\cite{Karananas:2024xja,Gialamas:2024iyu}\footnote{The inverse relations are
$$\Phi=\sqrt{6}m_1\sinh^{-1}(\phi/\sqrt{6} m_1)\,, \quad Z=-m_1\ln\left(\frac{\bar{\zeta}_1m_1}{2(\phi^2+6m_1^2)}\right)\,.$$}
\begin{equation}
 \phi =\sqrt{6}m_1 \sinh\left( {\frac{\Phi}{\sqrt{6}m_1}}\right) \,, \quad \bar{\z}_1 = 12m_1^2 e^{-Z/m_1}\cosh^2\left({\frac{\Phi}{\sqrt{6}m_1}} \right)\,, 
\end{equation}
which transforms $\mathcal{S}_1$ into
\begin{equation}
\mathcal{S}_1 = \int {\rm d}^4x \sqrt{-g_1} \left[ \frac{m_1^2}{2} R[g_1] -\frac12 (\partial_\m \Phi)^2 - \frac{1}{2}K_Z(\Phi,Z) (\partial_\m Z)^2 - V_1(\Phi,Z) \right]\,,
\end{equation}
where
\begin{subequations}
\label{eq:KzV1}
\begin{align}
K_Z(\Phi,Z) =& \frac{3}{2}\frac{\cosh^4\left(\frac{\Phi}{\sqrt{6}m_1}\right)}{\left[\cosh^2\left(\frac{\Phi}{\sqrt{6}m_1}\right)+\frac{e^{2Z/m_1}}{(24)^2}\left(1+288 C\sinh^2\left(\frac{\Phi}{\sqrt{6}m_1}\right)\right)\right]}\,,
\\[0.3cm]
V_1(\Phi,Z) =& \frac{m_1^4 f_1^2}{4} + 36m_1^4\tilde{f}_1^2e^{-2Z/m_1}\cosh^4\left(\frac{\Phi}{\sqrt{6}m_1}\right)\,.
\end{align}
\end{subequations}   

Before proceeding with the diagonalization of $\mathcal{S}_2$ we first perform a Weyl rescaling of the metric $g_{2\,\m\n}$, given by\footnote{Note that Weyl-invariance has been
used to gauge-fix the auxiliary $\chi_1$, so that no Weyl-rescaling for $g_{1\,\m\n}$ is required. In contrast, the metric $g_{2\,\m\n}$ enters in the Jordan frame and a Weyl-rescaling is appropriate.} 
\begin{equation}
 g_{2\,\m\n}\,\rightarrow\, \frac{m_2^2}{\bar{\chi}_2}g_{2\,\m\n}\,.
\end{equation}
The rescaling introduces the standard additional derivative term, $-\frac{3}{4}m_2^2 \frac{(\partial_\m\bar{\chi}_2)^2}{ \bar{\chi_2}^2}$. Consequently, the actions $\mathcal{S}_2$ and $\mathcal{S}_{\rm int}$ are modified accordingly:

\begin{align}
\label{eq:S2_resc}
&\mathcal{S}_2 = \int {\rm d}^4x \sqrt{-g_2} \bigg[  \frac{m_2^2}{2} R[g_2] -\frac{3}{4}m_2^2 \frac{(\partial_\m\bar{\chi}_2)^2}{\bar{\chi_2}^2} +\frac{3}{4}m_2^2 \frac{ (\partial_\m \bar{\chi}_2)^2-4 (\partial_\m\bar{\z}_2)^2+8 \frac{\bar{\z}_2}{\bar{\chi}_2}\partial_\m\bar{\chi}_2 \partial^\m\bar{\z}_2}{\bar{\chi}_2^2 +4\bar{\z}_2^2}\,, \nonumber
\\
& \hspace{3.2cm}  - \frac{1}{4}m_2^4f_2^2 - \frac{1}{4}m_2^4\tilde{f}_2^2\frac{\bar{\z}_2^2}{\bar{\chi}_2^2} \bigg]\,,
\\[0.2cm]
&\mathcal{S}_{\rm int} =  36m_1^4\int {\rm d}^4x \sqrt{-g_1}  \sinh^4\left( {\frac{\Phi}{\sqrt{6}m_1}}\right)\sum_{n=0}^{4} \left(\frac{m_2}{\sqrt{\bar{\chi}_2}} \right)^n\b_n e_n(\sqrt{\Delta})\,, 
\label{eq:Sint_resc}
\end{align}  
where we made use of the property $e_n(\sqrt{\lambda x})=\l^{n/2} e_n(\sqrt{x})$. Introducing the field
\begin{equation}
    \s = m_2 \frac{\bar{\z}_2}{\bar{\chi}_2}\,,
\end{equation}
${\cal{S}}_2$ reduces to a single-field action
\begin{equation}
\mathcal{S}_2 = \int {\rm d}^4x \sqrt{-g_2} \left[  \frac{m_2^2}{2} R[g_2] -\frac{1}{2}K_\s(\s)(\partial_\m\s)^2 - V_2(\s) \right]\,,
\end{equation}
where
\begin{equation}
\label{eq:KsV2}
K_\s(\s) = \frac{6}{1+4\s^2/m_2^2} \quad \text{and} \quad V_2(\s) =  \frac{1}{4}m_2^4 f_2^2 + \frac{1}{4}m_2^2 \tilde{f}_2^2\s^2\,.
\end{equation}

Our complete final action is given by
\begin{align}
\label{eq:act_full}
 \mathcal{S}_1 +  \mathcal{S}_2 +  \mathcal{S}_{\rm int} &=  \int {\rm d}^4x \sqrt{-g_1} \left[ \frac{m_1^2}{2} R[g_1] -\frac12 (\partial_\m \Phi)^2 - \frac{1}{2}K_Z(\Phi,Z)(\partial_\m Z)^2 - V_1(\Phi,Z) \right]  \nonumber
 \\
 & + \int {\rm d}^4x \sqrt{-g_2} \left[  \frac{m_2^2}{2} R[g_2] -\frac{1}{2}K_\s(\s)(\partial_\m\s)^2 - V_2(\s) \right] \nonumber
\\
& + 36m_1^4\int {\rm d}^4x \sqrt{-g_1}  \sinh^4\left( {\frac{\Phi}{\sqrt{6}m_1}}\right)\sum_{n=0}^{4} \left(\frac{m_2}{ \sqrt{\bar{\chi}_2}} \right)^n\b_n e_n(\sqrt{\Delta})\,,
\end{align}
where the involved functions are given by~\eqref{eq:KzV1} and \eqref{eq:KsV2}. 
The absence of ghost degrees of freedom in standard massive gravity~\cite{deRham:2010kj} and bigravity~\cite{Hassan:2011zd} was explicitly demonstrated in~\cite{Hassan:2011hr,Hassan:2011ea}. For actions similar to~\eqref{eq:act_full}, where the interaction potential is field-dependent, this was proven in~\cite{Gialamas:2023fly} and was also adopted in the context of multi-gravity in~\cite{Wood:2025mmq}.

Note that this action, apart from the two spin-2 particles, describes one dynamical scalar field and two dynamical pseudoscalars, namely $Z$ and $\sigma$, associated with the presence of the Holst terms. The appearing auxiliary $\chi_2$ is clearly non-dynamical, not possessing a kinetic term. As it has been pointed out, the coupling of the pseudoscalar sector to the axial vector current of SM fermions can lead to a possible solution of the strong CP problem~\cite{Karananas:2024xja}.

\subsection{Equations of motion}
\label{sec:3a}

Varying the action~\eqref{eq:Sint_resc} with respect to the non-dynamical scalar $\bar{\chi}_2$ we get
\begin{equation}
\label{eq:consteq}
    \frac{\d \mathcal{S}_{\rm int}}{\d \bar{\chi}_2} = 0\,\, \Rightarrow \,\, \sum_{n=0}^{4} \left( \bar{\chi}_2 \right)^{{-n/2}-1} n\mathring{\b}_n e_n(\sqrt{\Delta}) = 0\,, \quad \text{where} \quad \mathring{\b}_n = m_2^n \b_n\,.
\end{equation}
This equation amounts 
\begin{equation}
\label{eq:constr}
  \mathring{\b}_1e_1(\sqrt{\Delta})  \bar{\chi}_2^{3/2} +2\mathring{\b}_2e_2(\sqrt{\Delta})\bar{\chi}_2+3\mathring{\b}_3 e_3(\sqrt{\Delta}) \bar{\chi}_2^{1/2} +4\mathring{\b}_4e_4(\sqrt{\Delta})= 0\,,
\end{equation}
with the $\bar{\chi}_2$ coefficients becoming constant for proportional backgrounds. Note that for the so-called {\textit{minimal choice}}~\cite{Hassan:2011zd} ($\b_2=\b_3=0$),~\eqref{eq:constr} yields
\begin{equation}
\bar{\chi}_2^{1/2} = m_2 \left(\frac{-4\b_4e_4(\sqrt{\Delta})}{\b_1e_1(\sqrt{\Delta})} \right)^{1/3} \,.
\end{equation}

By varying the action~\eqref{eq:act_full} with respect to $\Phi, Z$ and $\s$, we obtain the scalar field equations of motion:
\begin{subequations}
\label{eq:scalar_eoms}
\begin{align}
&\Box \Phi -\frac12 \frac{\partial K_Z}{\partial \Phi} (\partial_\m Z)^2 +12\sqrt{6} m_1^3\sinh\left( \frac{2\Phi}{\sqrt{6}m_1}\right) 
\\   
& \hspace{0.6cm} \times\left[\sinh^2\left(\frac{\Phi}{\sqrt{6}m_1}\right)\sum_{n=0}^{4} \bar{\chi}_2^{-n/2} \mathring{\b}_n e_n(\sqrt{\Delta}) - \tilde{f}_1^2\cosh^2\left(\frac{\Phi}{\sqrt{6}m_1}\right) e^{-2Z/m_1} \right] = 0\,, \nonumber 
\\[0.2cm]
& \Box Z  -\frac12 \frac{\partial K_Z}{\partial Z} (\partial_\m Z)^2 +72m_1^3 \tilde{f}_1^2 e^{-2Z/m_1}\cosh^4\left(\frac{\Phi} {\sqrt{6}m_1}\right) = 0\,,
\\[0.2cm]
& \Box \s  -\frac12 \frac{\partial K_\s}{\partial \s} (\partial_\m \s)^2 - \frac{1}{2}m_2^2 \tilde{f}_2^2 \s = 0\,.
\end{align}
\end{subequations}
Variations with respect to the metrics $g_{1\,\m\n}$ and $g_{2\,\m\n}$ give the Einstein equations
\begin{subequations}
\begin{align}
\label{eq:Eins_a}
  R_{ \mu\nu}(g_1)-\frac{1}{2}g_{1\, \mu\nu}R(g_1) 
    & = {\frac{1}{m_1^2}}T^{(1)}_{\mu\nu} \,, 
    \\
    R_{ \mu\nu}(g_2)-\frac{1}{2}g_{ 2\,\mu\nu}R(g_2) 
    & =  {\frac{1}{m_2^2}}T^{(2)}_{ \mu\nu}\,, 
    \label{eq:Eins_b}
\end{align}
\end{subequations}
where
\begin{subequations}
\label{eq:Tmn_0}
\begin{align}
T^{(1)}_{\mu\nu} =&  \partial_\m\Phi\partial_\n\Phi -\frac12 g_{1\,\m\n}(\partial_\r\Phi)^2 +K_Z \partial_\m Z\partial_\n Z -\frac{K_Z}{2} g_{1\m\n} (\partial_\r Z)^2 - g_{1\,\m\n} V_1(\Phi,Z) \nonumber
\\
& -72m_1^4 \sinh^4\left(\frac{\Phi}{\sqrt{6}m_1}\right) \sum_{n=0}^{4} \mathring{\b}_n \frac{1}{\sqrt{-g_1}} \frac{\d}{\d g_1^{\,\m\n}}\left((\sqrt{-g_1} \bar{\chi}_2^{{-n/2}} e_n(\sqrt{\Delta} )\right)\,,
\\[0.2cm]
T^{(2)}_{\mu\nu} =& K_\s \partial_\m \s\partial_\n \s -\frac{K_s}{2} g_{2\,\m\n} (\partial_\r \s)^2 - g_{2\,\m\n} V_2(\s) \nonumber
\\
& -72m_1^4 \sinh^4\left(\frac{\Phi}{\sqrt{6}m_1}\right) \sum_{n=0}^{4} \mathring{\b}_n \frac{1}{\sqrt{-g_2}} \frac{\d}{\d g_2^{\,\m\n}}\left(\sqrt{-g_1} \bar{\chi}_2^{{ -n/2}} e_n(\sqrt{\Delta})\right)\,.
\end{align}
\end{subequations}
The remaining variations in the equations above, which have not yet been computed, can be expressed as
\begin{align}
\label{eq:vars}
 \sum_{n=0}^{4} \mathring{\b}_n  \frac{\d(\sqrt{-g_1} \bar{\chi}_2^{{-n/2}} e_n(\sqrt{\Delta} )}{\d g_a^{\,\m\n}} &= \sum_{n=0}^{4} \mathring{\b}_n \left[ \bar{\chi}_2^{{-n/2}} \frac{\d(\sqrt{-g_1}e_n(\sqrt{\Delta}))}{\d g_a^{\,\m\n}} - {\frac{n}{2}} \sqrt{-g_1}e_n(\sqrt{\Delta})  \bar{\chi}_2^{{-n/2 -1}} \frac{\d \bar{\chi}_2}{\d g_a^{\,\m\n}} \right] \nonumber
 \\
 & = \sum_{n=0}^{4} \mathring{\b}_n  \bar{\chi}_2^{{-n/2}} \frac{\d(\sqrt{-g_1}e_n(\sqrt{\Delta}))}{\d g_a^{\,\m\n}}\,,
\end{align}
where the second term on the right-hand side of the first line vanishes on-shell due to the equation of motion for $\bar{\chi}_2$~\eqref{eq:consteq}. At this stage, equation~\eqref{eq:vars} contains only the standard variations in the Bigravity framework, given by
\begin{equation}
\frac{-2}{\sqrt{-g_1}}\frac{\d(\sqrt{-g_1}e_n(\sqrt{\Delta}))}{\delta g_1^{\,\m\n}} = V_{\m\n}^{(n)} \quad \text{and} \quad
 \frac{-2}{\sqrt{-g_2}}\frac{\d(\sqrt{-g_1}e_n(\sqrt{\Delta}))}{\delta g_2^{\,\m\n}} = \tilde{V}_{\mu\nu}^{(4-n)}\,,    
\end{equation}
where
\begin{equation}
\label{eq:tensors_V}
V_{\m\n}^{(n)} = g_{1\,\m\l} Y^\l_{(n)\n}(\sqrt{g_1^{-1}g_2} ) \quad \text{and} \quad
\tilde{V}_{\mu\nu}^{(n)} = g_{2\,\m\l} Y^\l_{(n)\n}(\sqrt{g_2^{-1}g_1} )\,,
\end{equation}
with
\begin{equation}
Y_{(n)}(X) = \sum_{r=0}^n (-X)^{n-r}e_r(X)\,.
\end{equation}
As a result the energy momentum tensors~\eqref{eq:Tmn_0} can be written as
\begin{subequations}
\label{eq:Tmn_1}
\begin{align}
T^{(1)}_{\mu\nu} =&  \partial_\m\Phi\partial_\n\Phi -\frac12 g_{1\,\m\n}(\partial_\r\Phi)^2 +K_Z \partial_\m Z\partial_\n Z -\frac{K_Z}{2} g_{1\,\m\n} (\partial_\r Z)^2 - g_{1\,\m\n} V_1(\Phi,Z) \nonumber
\\
& +36m_1^4 \sinh^4\left(\frac{\Phi}{\sqrt{6}m_1}\right) \sum_{n=0}^{3} \mathring{\b}_n \bar{\chi}_2^{{-n/2}} V^{(n)}_{\m\n}\,,
\\[0.2cm]
T^{(2)}_{\mu\nu} =& K_\s \partial_\m \s\partial_\n \s -\frac{K_s}{2} g_{2\,\m\n} (\partial_\r \s)^2 - g_{2\,\m\n} V_2(\s) +36m_1^4 \sinh^4\left(\frac{\Phi}{\sqrt{6}m_1}\right) \sum_{n=1}^{4} \mathring{\b}_n \bar{\chi}_2^{{-n/2}} \tilde{V}^{(4-n)}_{\m\n}\,.
\end{align}
\end{subequations}

Closing this section we see that the equations of motion for the two gravity related pseudoscalars $\sigma$ and $\zeta_1$ are satisfied with $\sigma=0$ and $\zeta_1=0$. The latter corresponds to $Z=+\infty$ through the transformation relation $Z=-m_1\ln\left(\bar{\zeta}_1m_1/2(\phi^2+m_1^2)\right)$. A vanishing classical vacuum is also a solution for $\phi$. Nevertheless, in our general framework of consideration, where the effective theory of gravity is treated classically, while matter fields are sensitive to quantum corrections, it would be reasonable to expect that quantum corrections involving $\phi$ could be important in the flat limit. The potential expressed in terms of $\phi$ is just a quartic potential with the role of the bare coupling played by $\beta_0$, while the derivative interactions are suppressed for $\phi \ll m_1$. Note that $\phi$ could very well be a SM field or even the Higgs itself, interacting with the rest of matter. Dimensional transmutation via logarithmic corrections to the potential $\Delta V\,\sim\,\phi^4\ln(\phi^2/\mu^2)$ could take place, thus, generating dynamically a scale $\langle\phi\rangle=v\neq 0$. We are going to make this assumption in the rest our considerations of the above model.

\section{Background solutions and particle spectrum}
\label{sec:4}
In this section, we shall consider a special class of solutions for the metric tensors, known as proportional solutions. Perturbing around these solutions we derive the corresponding particle spectrum which includes the standard graviton and a massive spin-2 particle.

\subsection{Proportional solutions}
\label{sec:4a}
Proportional solutions~\cite{Hassan:2012wr} by definition are those for which the two metrics are related by a constant proportionality factor
\be
\label{eq:prop_back}
g_{2\,\m\n} = c^2 g_{1\,\m\n} = c^2 g_{\m\n}\,.
\ee
These solutions play a crucial role in simplifying the structure of the field equations\footnote{Note that for proportional backgrounds 
\be
\label{eq:vareps}
e_n(\sqrt{\Delta}) =\binom{4}{n}c^n\,.
\ee} and allow for analytic progress in understanding the vacuum structure and perturbative stability of the theory. Note that the requirement that both metrics should have a flat limit is easier to satisfy in the case of proportional metrics, in contrast to a case of independent metrics, which poses a more difficult issue to be studied.
Given this relationship, the tensors $V^{(n)}_{\m\n}$ and $\tilde{V}^{(n)}_{\m\n}$,  defined in~\eqref{eq:tensors_V}, take the form
\begin{equation}
     V_{\mu\nu}^{(n)} =g_{\m\n}\binom{3}{n}c^n \quad \text{and} \quad \tilde{V}_{\mu\nu}^{(n)}   =g_{\m\n}\binom{3}{n}c^{2-n}\,, \quad \text{with}\quad n=0,1,2,3\,.
\end{equation}
These expressions show that, under proportional solutions, the interaction tensors retain a simple dependence on $c$, greatly simplifying the field equations.

These solutions can be valid in vacuum states, where the fields acquire their vacuum expectation values, or along specific directions in field space~\cite{Gialamas:2023lxj}. In the following, we analyze the behavior of these solutions in the vacuum.
The equations of motion for the pseudoscalars are satisfied for $\s =0$ and $\zeta_1=0$, while, as we have argued in the previous section, the scalar $\phi$ can in principle acquire a non-vanishing expectation value.
The potential for $\Phi$, expressed in terms of the original variable $\phi$, arising only from the interaction term, for proportional backgrounds $g_{2\, \mu\nu}=c^2g_{1\, \mu\nu}$ corresponds to a quartic potential
\be 
V_{\rm int}=\phi^4\sum_{n=0}\mathring{\beta}_n\bar{\chi}_2^{{-n/2}}e_n(\sqrt{\Delta})\,\xRightarrow{\eqref{eq:prop_back}}\, V_{\rm int} = \phi^4\sum_{n=0}^{4}\left(\bear{c}
4\\
n
\eear\right)\mathring{\beta}_n\bar{\chi}_2^{-n/2}c^n\,,
\ee
where $\bar{\chi}_2$ is a constant solution of the equation $\sum_{n=1}^{4}\left(\bear{c}
4\\
n
\eear\right)n\mathring{\beta}_nc^n\bar{\chi}_2^{-n/2}=0$.
 Quantum corrections due to the $\phi$ self-interaction and to interactions with matter fields, will necessarily generate logarithmic corrections $\Delta V\propto \phi^4\ln({\phi^2}/{\mu^2})$, which are expected to break its classical scale invariance and lead to a non-vanishing expectation value $\langle\Phi\rangle=v_{\Phi}\neq 0$.

Assuming proportional backgrounds~\eqref{eq:prop_back} the energy momentum tensors at the vacuum read
\begin{subequations}
\label{eq:Tmn_2}
\begin{align}
T^{(1)}_{\mu\nu} =&  - g_{\m\n} \frac{1}{4}m_1^4 f_1^2+36m_1^4 \sinh^4\left(\frac{v_\Phi}{\sqrt{6}m_1}\right) \sum_{n=0}^{3} \mathring{\b}_n \bar{\chi}_2^{{-n/2}}\binom{3}{n}c^ng_{\m\n} = -m_1^2 \Lambda_1g_{\m\n}\,,
\\[0.2cm]
T^{(2)}_{\mu\nu} =&  - g_{\m\n} \frac{1}{4}c^2m_2^4 f_2^2 +36m_1^4 \sinh^4\left(\frac{v_\Phi}{\sqrt{6}m_1}\right) \sum_{n=1}^{4} \mathring{\b}_n \bar{\chi}_2^{{-n/2}} \binom{3}{4-n}c^{n-2}g_{\m\n} \nonumber
\\
 =& - g_{\m\n} \frac{1}{4}c^2m_2^4 f_2^2 = -\a^2m_1^2 \Lambda_2g_{\m\n} \,,
\end{align}
\end{subequations}
where $\a = m_2/m_1$ represents the ratio of the gravitational masses in the two sectors. Note that the $\mathring{\b}_n$-dependent term of $T^{(2)}_{\mu\nu}$ vanishes on-shell due to the equation of motion ({\ref{eq:consteq}}) for $\bar{\chi}_2$. This follows from the combinatorial identity $\binom{3}{4-n} = \frac{n}{4} \binom{4}{n}$.

A crucial property of the Einstein tensor, $R_{\m\n} - \frac{1}{2}g_{\m\n} R$, is that it remains invariant under uniform scaling of the metric as in~\eqref{eq:prop_back}.  Consequently, the left-hand sides of the Einstein equations~\eqref{eq:Eins_a} and~\eqref{eq:Eins_b} are identical, leading to the proportionality of their corresponding energy-momentum tensors, that is $\a^2 T^{(1)}_{\mu\nu} = T^{(2)}_{\mu\nu}$.
By redefining the parameters $\mathring{\b}_n$ as $\bar{\b}_n = \mathring{\b}_n {c^n \bar{\chi}_2^{-n/2}}$ and using the proportionality of the energy-momentum tensors, we obtain the following parameter relation
\begin{equation}
\frac{1}{4}f_1^2 -36\sinh^4 \left(\frac{v_\Phi}{\sqrt{6}m_1}\right) (\bar{\b}_0+3\bar{\b}_1+3\bar{\b}_2+\bar{\b}_3)  = \frac{1}{4}\bar{\a}^2f_2^2\,,
\end{equation}
where $\bar{\a}=\a c$. 
Furthermore, the proportionality condition affects the equation of motion for
$\bar{\chi}_2$, given by~\eqref{eq:constr}, which takes the form $ \mathring{\b}_1 \bar{\chi}_2^{3/2} + 3c\mathring{\b}_2 \bar{\chi}_2 +3c^2\mathring{\b}_3 \bar{\chi}_2^{1/2} +c^3\mathring{\b}_4 = 0  $. This is a cubic equation for $\bar{\chi}_2^{1/2}$ that can be solved to express $\bar{\chi}_2$ as a function of the parameters, that is, $ \bar{\chi}_2  = \bar{\chi}_2  (\bar{\a},\b_1,\b_2,\b_3,\b_4)$.

\subsection{Linearization}
\label{sec:4b}
Perturbation theory in bimetric gravity faces significant challenges due to the presence of the square root matrix in the interaction potential. This complication generally makes the study of fluctuations more intricate. Nevertheless, for proportional solutions, the structure of the equations simplifies considerably, allowing for a more tractable analysis and providing a simplified setting for analyzing fluctuations. 

Linear perturbations around proportional backgrounds are studied by introducing the metric fluctuations $h_{\m\n}$ and $l_{\m\n}$ as deviations from the background metrics $g_{1\,\m\n}$ and $g_{2\,\m\n}$ respectively. Specifically, we define
\be 
\label{eq:background} g_{1\,\mu\nu} = g_{\mu\nu} + h_{\mu\nu}\quad \text{and} \quad g_{2\,\mu\nu} = c^2 g_{\mu\nu} + l_{\mu\nu}\,. 
\ee
Since the theory generically leads to a massive and a massless spin-2 excitation, it is useful to construct linear combinations of $h_{\m\n}$ and $l_{\m\n}$ that diagonalize the mass spectrum. The canonically normalized mass eigenstates are then given by
\be
\label{eq:eigen}
G_{\m\n} = \frac{M_P}{(1+\a^2c^2)}\left(h_{\m\n}+\a^2l_{\m\n} \right)\,, \qquad M_{\m\n} = \frac{\a M_P}{c(1+\a^2c^2)}\left(l_{\m\n}-c^2h_{\m\n} \right)\,,
\ee
where $G_{\m\n}$ corresponds to the massless spin-2 mode and $M_{\m\n}$ represents the massive eigenstate. The physical Planck mass, which governs gravitational interactions, is defined as\footnote{We shall assume that the coupling to matter is realized exclusively through $g_{1\, \mu\nu}$. In this case 
$$T_{ \mu\nu}^{(m)}g_1^{\, \mu\nu} = T_{ \mu\nu}^{(m)}g^{ \mu\nu}+ \frac{1}{M_P}G^{ \mu\nu}T_{ \mu\nu}^{(m)}-\frac{\bar{\alpha}}{M_P}M^{ \mu\nu}T_{ \mu\nu}^{(m)}\,.$$} $M_P=\sqrt{m_1^2+c^2m_2^2}$. These redefinitions ensure that the kinetic terms for the mass eigenstates are properly normalized, simplifying the analysis of their interactions. The linearized action for these perturbations takes the form
\begin{align}
\label{eq:inter_action}
\mathcal{S} = &\frac12 \int {\rm d}^4x \sqrt{-g}  \bigg[\frac14 G_{\m\n} \mathcal{E}^{\m\n\r\s}G_{\r\s} + \frac14 M_{\m\n} \mathcal{E}^{\m\n\r\s}M_{\r\s} +\frac{\Lambda_1}{2} \left(G_{\m\n}G^{\m\n} -\frac12 G^2\right) \nonumber
\\
&  +\frac{\Lambda_1}{2} \left(M_{\m\n}M^{\m\n} -\frac12 M^2\right) -\frac{m_{\rm FP}^2}{4}\left(M_{\m\n}M^{\m\n} - M^2\right) - M_P \Lambda_1 G -2M_P^2 \Lambda_1 \bigg]\,,
\end{align}
where we have used the equality of cosmological constants\footnote{Note that $\Lambda_1=m_1^2\left( \frac{1}{4}f_1^2 -36\sinh^4 \left(\frac{v_\Phi}{\sqrt{6}m_1}\right) (\bar{\b}_0+3\bar{\b}_1+3\bar{\b}_2+\bar{\b}_3)  \right)$.} $\Lambda_1 =\Lambda_2$. The structure of this action highlights the presence of two independent sectors: one corresponding to a massless graviton and another to a massive spin-2 field. The massive sector exhibits a standard Fierz-Pauli structure with mass term $m_{\rm FP}^2$, which ensures that the theory propagates the correct number of degrees of freedom without introducing ghost-like instabilities. The Fierz-Pauli mass of the massive spin-2 field is explicitly given by
\begin{equation}
\label{eq:mfp}
    m_{\rm FP}^2 = -\frac{36M_P^2}{\bar{\a}
    ^2}\sinh^4\left(\frac{v_\Phi\sqrt{1+\bar{\a}^2}}{\sqrt{6}M_P}\right) ( \bar{\b}_1 +2\bar{\b}_2 +\bar{\b}_3)\,.
\end{equation}
This mass term arises from the interaction potential between the two metrics and plays a crucial role in determining the phenomenology of the massive spin-2 mode.

\section{Massive spin-2 phenomenology}
\label{sec:5}
In this section, we shall examine the Fierz-Pauli mass~\eqref{eq:mfp} and see how it relates to the model parameters, specifically $v_\Phi$, the ratio of the gravitational scales  $\bar{\a}$, and the $\b$ parameters.  We shall also investigate whether the massive spin-2 particle can constitute DM and, using previously established constraints, derive limits on the vacuum expectation value $v_{\Phi}$.
We shall focus on the so-called minimal choice of parameters, where the parameters $\b_2$ and $\b_3$ are set to zero. This choice simplifies the analysis while preserving the essential features of the theory. In this scenario, solving the equation of motion for $ \bar{\chi}_2$, we obtain
\begin{equation}
 \bar{\chi}_2^{1/2} = \frac{\bar{\a}M_P}{\sqrt{1+\bar{\a}^2}} \left(-\b_4/\b_1\right)^{1/3}\,,  
\end{equation}
which leads to the rescaled $\bar{\b}$ parameters
\begin{equation}
\bar{\b}_1 = \frac{\b_1}{(-\b_4/\b_1)^{1/3}} \quad \text{and} \quad \bar{\b}_4 = \frac{\b_4}{(-\b_4/\b_1)^{4/3}} \,.
\end{equation}
 Note that even in the most complex case, where none of the $\beta$ parameters are zero, the $\bar{\b}$ parameters always depend solely on the $\beta$ parameters and not on any other parameters of our model.
To explore the behavior of the Fierz-Pauli mass, we take the $\b$ parameters to be of order $\mathcal{O}(1)$, and analyze how the mass evolves for different values of the vev $v_\Phi$, ranging between $\mathcal{O}(M_W)$ and $\mathcal{O}(M_P)$. Taking the specific values $\b_1=-1=-\b_4$, the Fierz-Pauli mass becomes
\begin{figure}[t]
\begin{center}
\includegraphics[width=0.485\textwidth]{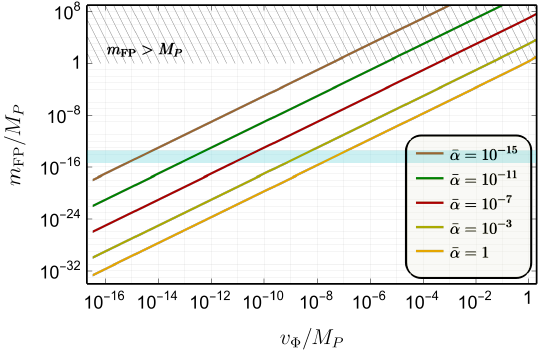} \hspace{0.2cm} \includegraphics[width=0.485\textwidth]{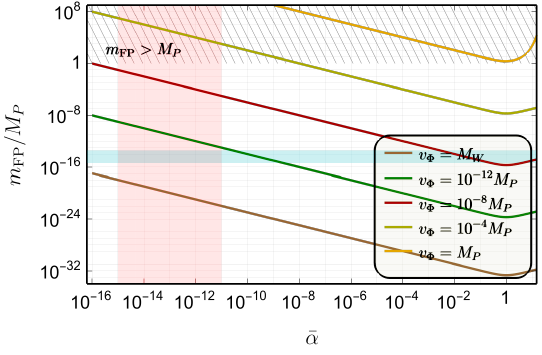}
\end{center}
\caption{The Fierz-Pauli mass, $m_{\rm FP}$, as function of the $v_\Phi$ (left) and $\bar{\a}$ (right),  with parameters $\b_1=-1=-\b_4$. The cyan and red shaded regions indicate the constraints on $m_{\rm FP}$ and $\bar{\a}$ derived in~\cite{Babichev:2016bxi}, under the assumption that the massive spin-2 particle constitutes DM.}
\label{fig:fig1}
\end{figure}
\begin{equation}
\label{eq:mfpav}
    m_{\rm FP} = \frac{6M_P}{\bar{\a}
    }\sinh^2\left(\frac{v_\Phi\sqrt{1+\bar{\a}^2}}{\sqrt{6}M_P}\right) \,.
\end{equation}  
In figure~\ref{fig:fig1} we show the Fierz-Pauli mass as function of $v_\Phi$ (left) and $\bar{\a}$ (right), computed using equation~\eqref{eq:mfpav}. It is clear that the Fierz–Pauli mass can vary significantly depending on the values of $v_\Phi$ and $\bar{\a}$. In the following, we discuss the limiting cases of ultralight and superheavy massive spin-2 particles as well as the case of the spin-2 particle being dark matter~\cite{Aoki:2014cla,Aoki:2016zgp,Babichev:2016hir,Babichev:2016bxi,Marzola:2017lbt,Manita:2022tkl,Kolb:2023dzp,Chen:2023oip,Cai:2024thd,Blas:2024jyh,Zhang:2025fck}.\\ 

\noindent\textit{1. \underline{Ultralight massive spin-2}}

% \noindent
For small vevs, i.e., $v_\Phi\ll M_P$, assuming $\bar{\alpha}<1$, the Fierz-Pauli mass formula simplifies to the approximate form $m_{\rm FP} \simeq v_\Phi^2/(\bar{\a}M_P)$, which shows that it scales quadratically with $v_\Phi$, suppressed by $\bar{\a}M_P$. If we take $v_\Phi$ to be of the order of the electroweak scale, $v_\Phi \sim M_W$, the corresponding mass is estimated to $m_{\rm FP} \simeq 3\times 10^{-6}\, \bar{\a}^{-1}$ eV, indicating an ultralight massive spin-2 particle in the limiting case where $\bar{\a}\sim \mathcal{O}(1)$. As $\bar{\a}$ decreases, the Fierz-Pauli mass correspondingly increases, highlighting its inverse dependence on $\bar{\a}$.\\

\noindent\textit{2. \underline{Superheavy massive spin-2}}

Restricting our analysis to the case where $\bar{\a}<1$ and considering a vev near the Planck scale, we find a Fierz-Pauli mass $m_{\rm FP}\sim M_P/\bar{\a}$, which is always super-Planckian\footnote{ See also the orange line in the right panel of figure~\ref{fig:fig1}.}. This excludes such high values for the vev $v_{\Phi}$. 
However, if we consider lower vev values $v_\Phi \sim 10^{-\n} M_P$, as for example at the GUT scale ($\nu\sim 2$ or $3$), we obtain $m_{\rm FP} \sim 10^{-2\n} M_P/\bar{\a}$.
For $m_{\rm FP}$ to remain sub-Planckian, we have to require $\bar{\a} > 10^{-2\n}$. More generally, since we cannot have masses exceeding the Planck scale, we must have $v_{\Phi}\lesssim M_P\sqrt{\bar{\alpha}}$.

The above distinction between the ultralight and superheavy regimes is based on our assumption that the $\b$ parameters are of order $\mathcal{O}(1)$. Since these parameters act as multiplicative factors in the Fierz-Pauli mass, they can significantly alter its value if we assume that $\b$ is either very large or very small. This would allow for a smooth transition between the ultralight and superheavy regimes by appropriately tuning the $\b$ parameters. In general, there is considerable flexibility in setting the mass of the spin-2 particle, allowing us to satisfy various constraints, such as those arising when the massive spin-2 particle is identified as a candidate for dark matter~\cite{Babichev:2016hir,Babichev:2016bxi}.\\

\noindent\textit{3. \underline{Massive spin-2 as dark matter}}

It has been well established that the massive spin-2 particles predicted by bimetric gravity could serve as viable DM candidates due to their extremely weak interactions with SM particles. The parameter space for massive spin-2 DM can be constrained by considering the most stringent lifetime bounds for decaying DM, limits imposed by DM overproduction, and the exclusion of regions where perturbative analysis is not applicable.
By incorporating these constraints, the authors of~\cite{Babichev:2016bxi} found that the allowed range for the Fierz-Pauli mass is significantly restricted to $1\, \text{TeV} \lesssim m_{\rm FP} \lesssim 100\, \text{TeV}$, while the gravitational mass associated with the second metric lies within the range is $1\, \text{TeV} \lesssim m_2\lesssim 10^4\, \text{TeV} $. These values, in turn, correspond to the parameter range $10^{-15} \lesssim \bar{\a} \lesssim 10^{-11}$.  The regions satisfying these constraints are indicated by the shaded areas in figure~\ref{fig:fig1}. Applying these bounds, the vacuum expectation value of the scalar field is determined to be
\begin{equation}
1.5\times 10^3 \, \text{GeV} \lesssim v_\Phi \lesssim 1.5\times 10^6 \, \text{GeV}\,.
\end{equation}
As mentioned earlier, this result holds under the assumption that the parameters $\b$ are of order $\mathcal{O}(1)$. If the 
$\b$ parameters take on smaller (larger) values, they lead to larger (smaller) vevs. Note that if we relax the assumption that the $\beta$'s are of $\mathcal{O}(1)$, the massive spin-$2$ particles can constitute DM with the characteristic vev of the SM Higgs boson $v_H\sim 246\,GeV$ for a wide range $10^3 \lesssim \b \lesssim 10^{15}$

To conclude this section, let's have a closer look at the arising cosmological constants. They are given by
\begin{subequations}
\begin{align}
    \Lambda_1 =&  \frac{f_1^2 M_P^2 }{4(1+\bar{\a}^2)} - \frac{36M_P^2}{1+\bar{\a}^2} \sinh^4\left(\frac{v_\Phi\sqrt{1+\bar{\a}^2}}{\sqrt{6}M_P}\right) \left(\b_0 -3 \right)\,,
\\ 
\Lambda_2 =& \frac{\bar{\a}^2 f_2^2 M_P^2}{4(1+\bar{\a}^2)}\,,
\end{align}
\end{subequations}
where we have used the relation $\bar{\b}_0 +3\bar{\b}_1+3\bar{\b}_2 +\bar{\b}_3 = \b_0-3$ for $\b_1=-1$. In order to satisfy the observational constraint $\Lambda_1 =\Lambda_2 \simeq 10^{-120}\, M_P^2$ we can set $\b_0 =3$ to cancel the dominant contribution to the cosmological constant, arising from the non-zero vev. Then, the gravitational contributions, which have to satisfy the constraint $f_1=\bar{\alpha}f_2$, can meet the observational constraint for exceedingly small values of the parameters $f_1 = \bar{\a}f_2 \sim 10^{-60}$. Nevertheless, additional perturbative and non-perturbative contributions should be expected that invalidate these arrangements. For instance, the contribution from the SM, mostly affecting $\Lambda_1$, is expected to be of $\mathcal{O}(M_W^4)$. Thus, a milder, equally agnostic, assumption could be made that the above gravitational contributions are at least of that order, leading to $f\sim \mathcal{O}(M_W^2/M_P^2)\approx 10^{-32}$. Note that, assuming that the couplings $\tilde{f}$ are of the same order as the $f$-couplings,  the smallness of the latter reflects directly on the  smallness of the axion $Z,\,\sigma$ masses, being $M_{1}\sim\,\frac{\tilde{f}_1M_P}{\sqrt{1+\bar{\alpha}^2}}$ and $M_{2}\sim\,\frac{\bar{\alpha}\tilde{f}_2M_P}{\sqrt{1+\bar{\alpha}^2}}$.

\section{Summary and Conclusions}
\label{sec:6}

 We started by considering the ghost-free bimetric theory of gravity, characterized by two independent metric tensors accompanied by two independent gravitational mass scales. Motivated by the use of classical scale invariance to face large scale hierarchies in particle physics, we imposed on bimetric gravity local Weyl invariance. We restricted the gravitational action to include at most quadratic terms, avoiding any unphysical degrees of freedom. The theory was coupled to a scalar field that modified the standard bigravity interaction potential by a quartic scalar field factor, thus, rendering it Weyl invariant. 

Furthermore, local Weyl invariance was realized in the framework of Einstein-Cartan gravity, with the role of a gauge vector taken up by a corresponding torsion vector and, thus, not introducing any additional dynamical degrees of freedom. The action, depending only on the dimensionless gravitational couplings and on the dimensionless parameters of the potential and consisting of quadratic terms of the Ricci scalars and the Holst invariants, was written in terms of axion-like pseudoscalars and two auxiliary scalar fields, the latter being exchanged for the two independent gravitational mass scales $m_1$ and $m_2$.

The scalar field, being massless at the classical level, is expected to receive radiative corrections either from its self-interaction or from possible interactions with the rest of matter. These corrections, unavoidably leading to the breaking of scale invariance at the quantum level, are likely to generate a non-vanishing expectation value through logarithms $\sim \phi^4\ln(\phi^2/\mu^2)$ in analogy to standard dimensional transmutation. Assuming this, we have considered metric perturbations around proportional backgrounds $g_{2\,\mu\nu}=c^2g_{1\, \mu\nu}$ and analyzed the resulting linearized theory. Consistency leads to a proportionality relation between the corresponding energy-momentum tensors that results in a relation among parameters. The spectrum of the theory, apart from the standard massless graviton, includes a pair of axion-like pseudoscalars, the scalar and the additional spin-$2$ state having the Fierz-Pauli mass. The latter depends on the vacuum expectation value of the scalar field, on the ratio $\bar{\alpha}=c(m_2/m_1)$ and on the dimensionless $\beta$ parameters of the interaction potential. 

Having established the general structure of the spectrum and the parametric constraints of the model, we proceeded to analyze the phenomenology of the model focusing on the massive spin-$2$ state. 
Assuming that the initial $\beta$ parameters of the standard bigravity potential are of order $\mathcal{O}(1)$, we found that the mass of the spin-2 state could range from ultralight values, $\mathcal{O}(10^{-6}\,{\rm eV})$, to superheavy approaching the Planck scale. Finally, the massive spin-2 particle can be identified as dark matter, provided that the vacuum expectation value of the scalar field lies within the range $\mathcal{O}(10^{3}\,{\rm GeV}) \lesssim v_\Phi \lesssim \mathcal{O}(10^{6}\,{\rm GeV})$, although this range could shift for larger or smaller values that include the SM Higgs vev if we relax the assumption that the $\beta$ parameters are of $\mathcal{O}(1)$.

As for future work, it would be interesting to perform a thorough analysis of cosmological backgrounds in our setup. In particular, it would be important to investigate the behavior of perturbations around cosmological solutions and assess their stability conditions~\cite{Volkov:2011an,Koennig:2014ods,Lagos:2014lca}, as well as to study the impact of the Fierz–Pauli mass on late-time cosmology, including possible implications for dark energy~\cite{Dwivedi:2024okk,Smirnov:2025yru,Hogas:2025ahb}.

%-------------------------------------------------------------------------------
\acknowledgments
%-------------------------------------------------------------------------------

IDG thanks Hardi Veerm\"ae  for useful discussions. The work of IDG was supported by the Estonian Research Council grants MOB3JD1202, RVTT3,  RVTT7, and by the CoE program TK202 ``Fundamental Universe''.

 \bibliography{bigravity_refs}
\end{document}